\documentclass[a4paper,twocolumn,11pt,accepted=2021-10-15,aps,superscriptaddress]{quantumarticle}
\pdfoutput=1
\usepackage[utf8]{inputenc}
\usepackage[english]{babel}
\usepackage[T1]{fontenc}
\usepackage{hyperref}
\usepackage{float}
\usepackage{amsfonts, amsmath, amssymb, bm, bbm, graphicx, epsfig, epstopdf}
\usepackage{braket}
\usepackage{textcomp}

\usepackage{color}
\definecolor{forestgreen}{RGB}{34,139,34}
\usepackage[normalem]{ulem} 
\usepackage{color} 

\usepackage{etoolbox}
\renewcommand{\doibase}[1]{https://doi.org/\ifdefempty{#1}{}{#1}}

\begin{document}
	\title{Quantum repeaters based on individual electron spins and nuclear-spin-ensemble memories in quantum dots}
	\author{K. Sharman}
	\affiliation{Institute for Quantum Science and Technology, and Department of Physics \& Astronomy, University of Calgary, 2500 University Drive NW, Calgary, Alberta T2N 1N4, Canada}
	\author{F. Kimiaee Asadi}
	\affiliation{Institute for Quantum Science and Technology, and Department of Physics \& Astronomy, University of Calgary, 2500 University Drive NW, Calgary, Alberta T2N 1N4, Canada}
	\author{S. C. Wein}
	\affiliation{Institute for Quantum Science and Technology, and Department of Physics \& Astronomy, University of Calgary, 2500 University Drive NW, Calgary, Alberta T2N 1N4, Canada}
	\author{C. Simon}
	\affiliation{Institute for Quantum Science and Technology, and Department of Physics \& Astronomy, University of Calgary, 2500 University Drive NW, Calgary, Alberta T2N 1N4, Canada}
	\begin{abstract}
	    Inspired by recent developments in the control and manipulation of quantum dot nuclear spins, which allow for the transfer of an electron spin state to the surrounding nuclear-spin ensemble for storage, we propose a quantum repeater scheme that combines individual quantum dot electron spins and nuclear-spin ensembles, which serve as spin-photon interfaces and quantum memories respectively. We consider the use of low-strain quantum dots embedded in high-cooperativity optical microcavities. Quantum dot nuclear-spin ensembles allow for the long-term storage of entangled states, and heralded entanglement swapping is performed using cavity-assisted gates. We highlight the advances in quantum dot technologies required to realize our quantum repeater scheme which promises the establishment of high-fidelity entanglement over long distances with a distribution rate exceeding that of the direct transmission of photons.
	\end{abstract}

\maketitle
\section{Introduction}
\label{Intro}

The establishment of a quantum internet \cite{kimble2008quantum, simon2017towards, wehner2018quantum} is required to realize many promising applications of quantum science including long-distance quantum key distribution \cite{bb84}, dense coding \cite{dense_coding, superdense_coding}, and distributed quantum computing \cite{distributed_comp}. Photons are a valuable resource for encoding and transmitting quantum information because they move quickly, exhibit quantum behavior at room temperature, and can be conveniently initialized and manipulated using linear optical devices. The transmission of photons using typical telecommunication optical fibers is, however, subject to loss and the rate of photon transmission decreases exponentially with the length of the communication channel. Classical telecommunications overcomes this loss through the use of optical amplifiers. A similar approach is not possible in quantum communication due to the no-cloning theorem \cite{no_cloning}, which states that it is impossible to create a copy of an unknown quantum state. This problem can be overcome in quantum communication using the quantum repeater approach \cite{first_proposal}.

The most common approach to quantum repeaters is to distribute entanglement over the length of a communication channel via a series of locally established links \cite{atomic_ensembles}. Entanglement is first generated over short distances, referred to as local links, and then the entanglement is stored in quantum memories. The local links are connected through entanglement swapping \cite{zukowski1993event} to extend the entanglement over longer distances. The result of the protocol is the generation of an entangled state between two qubits, referred to as communication qubits, which can be used to teleport quantum states over the length of the communication channel. Quantum memories \cite{simon2010quantum, heshami2016quantum} eliminate the need for simultaneous entanglement generation in each of the local links which is difficult to achieve due to unavoidable optical losses.

A variety of platforms have been proposed for the construction of a quantum repeater, including atomic-ensembles \cite{DLCZ},  individual rare-earth ions \cite{asadi2018quantum, asadi2020protocols}, nitrogen-vacancy centers in diamond \cite{Childress}, and semiconductor quantum dots \cite{Simon_qds}.

Single electron spins confined in quantum dots (QDs) offer fast initialization times, optical manipulation, and provide a promising route towards scalable quantum devices with a large number of qubits. QDs have been established as reliable emitters of single photons with near-unity indistinguishability \cite{Single_photon_sources, ding2016demand, wang2019towards} and there exist deterministic and reproducible methods to fabricate QD-cavity devices \cite{ollivier2020reproducibility}. Experimental work has already succeeded in producing entanglement between a QD electron and a single photon \cite{Observ_spin_photon, schaibley2013demonstration, de2012quantum}, as well as the heralded entanglement between distant QD electrons \cite{stockill2017phase, Heralded_entanglement_bw_hole_spins}.

In the past decade there has been significant progress in the ability to control the interaction between light and matter through the use of QDs embedded in photonic nanostructures \cite{lodahl2015interfacing}. Strong coupling has been demonstrated in a single QD-microcavity system \cite{reithmaier2004strong}, and probabilities greater than $98\%$ for photon emission into a desired waveguide mode have been reported for QDs coupled to a photonic crystal waveguide \cite{arcari2014near}. Site-controlled QD fabrication techniques such as the pyramid, inverted pyramid, nanohole, and spatially selective H incorporation techniques \cite{pettinari2018site} demonstrate that the near-deterministic positioning of QDs within nanostructures is possible. Continued improvements in semiconductor fabrication technologies are quickly advancing the fields of scalable, on-chip sources of quantum light \cite{on_chip_phcs}, and there have already been numerous efforts to integrate QDs with photonic circuits \cite{davanco2017heterogeneous, zadeh2016deterministic}.

A potential issue when considering the use of QD electron-spin qubits arises from the hyperfine interaction between the electron spin and QD nuclear spins. Nuclear spin diffusion occurs both through the nuclear dipole-dipole interaction and through a long-ranged nuclear spin interaction in which nuclear spin flip-flops can occur virtually through the electron spin \cite{gong2011dynamics}. The spin exchange flip-flops result in a fluctuating magnetic field (the Overhauser field) experienced by the electron spin \cite{chekhovich2015suppression}. The noisy nuclear spin environment contributes to the dephasing of the electron spin and has also limited the proposal of promising quantum memories.

If QD electron spin states are used as communication qubits, then there are two approaches for dealing with the influence of the nuclear ensemble. The first is to mitigate the unwanted effects via decoupling techniques, or possibly by using QDs fabricated from isotopically purified II-VI materials \cite{Simon_qds}. These methods could extend the coherence time of the electron spins, such that it is natural to ask if they could be used as both communication qubits and quantum memories.

Quantum repeaters are expected to outperform the direct transmission of photons for distances greater than $\sim 500$ km \cite{asadi2018quantum, atomic_ensembles}. For a memory-based quantum repeater to be competitive, we require quantum memories with coherence times ($T_2$) that are an order of magnitude larger than the entanglement generation time (see Sec. \ref{Rates}). The coherence times offered by QD electron spins \cite{gillard2021fundamental} are not likely to be sufficient to outperform direct transmission for distances greater than $\sim 500$ km, and as such, electron spins alone are not a viable option for QD-based quantum repeaters. 

The second approach is to consider the hyperfine interaction as a valuable resource, which can be used to control the nuclear spins and in fact turn the ensemble into a quantum memory. QD nuclear spins have the potential to serve as memories with exceptionally long storage times, as lifetimes of several hundred seconds have already been demonstrated for QDs \cite{ulhaq2016vanishing} suggesting that coherence times on the order of seconds are within reach (see Sec. \ref{initialization_sec}). Nuclear-ensemble memories would, by default, associate every QD electron spin qubit with an intrinsic local memory, resulting in a scalable platform. The notion of long storage times in such a scalable system has led to several proposals for quantum memories based on QD nuclear spin ensembles \cite{Long_lived_memory, controlling_mesoscopic_spin_envir, taylor2004quantum}.

Recent work \cite{Mapping} has experimentally demonstrated that the spin state of a QD electron can be mapped to the surrounding QD nuclear-spin ensemble. The work of Gangloff \textit{et al.} \cite{Mapping} has provided the steps necessary to realize nuclear-spin-ensemble-based quantum memories. Here we propose a quantum repeater protocol which utilizes QD electron-spin qubits and nuclear-spin-ensemble memories as communication and memory qubits, respectively.

The paper is organized as follows. In Sec. \ref{Protocol} we introduce our quantum repeater proposal and discuss the underlying principles of each component in the scheme. The entanglement distribution rates are provided in Sec. \ref{Rates}. We present the fidelity calculations for each of the components of our scheme in Sec. \ref{Fidelity}, along with the overall fidelity. Considerations for the possible implementation of the proposal are discussed in Section \ref{Implementation}. We conclude in Sec. \ref{Conclusion}.

\section{Quantum repeater protocol}
\label{Protocol}

In this repeater proposal, each node of the system consists of two electron-spin qubits, both of which are confined in separate QDs and are magnetically coupled to the nuclei of the surrounding dot ($N$ spin-$I$ nuclei). In each node, both QDs are embedded in the same optical microcavity. The system operates at liquid helium temperature (4 K). We limit our analysis to a single nuclear spin species and provide justification of this assumption in the context of the nuclear-spin-ensemble memory (Sec. \ref{Subsection_mapping}).

We propose the use of nanohole-filled droplet epitaxial (NFDE) GaAs/AlGaAs QDs \cite{atkinson2012independent} in our repeater scheme. This growth method produces QDs with levels of strain that are sufficiently homogeneous to realize a high-fidelity state transfer (see Sec. \ref{Subsection_mapping}), yet the small variance in strain across the dot results in the suppression of nuclear spin diffusion \cite{ulhaq2016vanishing} leading to a stable nuclear-spin environment. 

Experimental findings and nanofabrication techniques for NFDE QDs are not as well-developed as those of self-assembled QDs grown using standard methods such as the Stranski–Krastanov method. Our analysis requires numerous parameters that have not yet been measured for NFDE QDs. As such, we draw inspiration from various self-assembled QD systems, clearly indicating when we do. A detailed discussion of the progress that needs to be made with NFDE GaAs QDs in order to realize our repeater protocol is presented in Sec. \ref{Implementation}.

\begin{figure}
	\centering
	\includegraphics[width=8 cm]{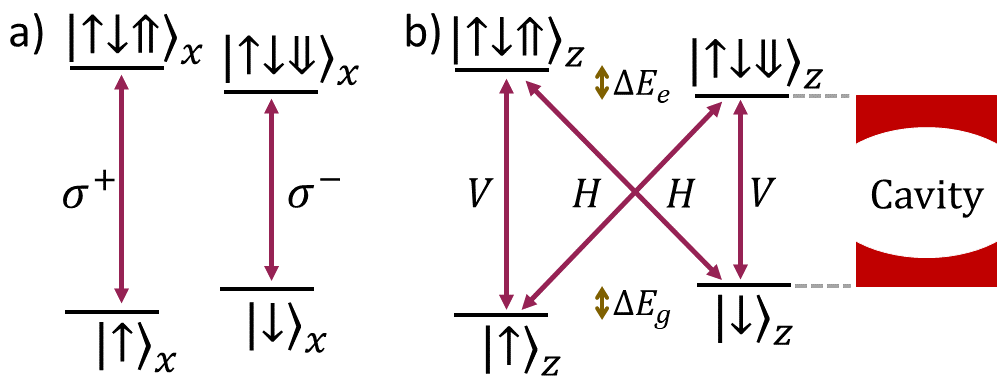}
	\caption{QD energy level diagram for (a) Faraday and (b) Voigt configurations of the applied magnetic field. $\uparrow$ ($\downarrow$) denotes an electron with spin up (down), $\Uparrow$ ($\Downarrow$) denotes a hole with spin up (down), and the subscript denotes the spin basis. The transitions are driven by circularly ($\sigma^\pm$), vertically ($V$), or horizontally ($H$) polarized light. Our scheme will consider the use of the Voigt configuration with excited (ground) states energy splitting $\Delta E_e$ ($\Delta E_g$) and a cavity coupled to the $\ket{\downarrow} \leftrightarrow \ket{\uparrow \downarrow \Downarrow}$ transition. The $\ket{\uparrow} \leftrightarrow \ket{\uparrow\downarrow\Uparrow}$ transition is far-detuned from the cavity.}\label{e_levels}
\end{figure}

The direction of an applied magnetic field with respect to the growth direction of the QD will dictate the allowable optical transitions. In the Faraday configuration, where the magnetic field is applied parallel to the QD growth direction (taken to be the $x$-direction), the electron spin eigenstates, denoted $\ket{\uparrow}_x$ and $\ket{\downarrow}_x$, are split by the Zeeman effect. The optically excited eigenstates are known as trion states, which consist of an electron spin singlet and a hole spin, denoted $\ket{\uparrow\downarrow\Uparrow}_x$ and $\ket{\uparrow\downarrow\Downarrow}_x$. There are two strongly allowed transitions that are driven by circularly polarized light ($\sigma^{\pm}$), as shown in Fig. \ref{e_levels}a.

In the Voigt configuration, where the magnetic field is applied in the growth plane ($z$-direction), there are four transitions of equal strength that are driven by linearly polarized light. The vertical transitions are driven with vertically-polarized light ($V$), and the diagonal transitions with horizontally polarized light ($H$), as shown in Fig. \ref{e_levels}b.

The repeater protocol described here will make use of the Voigt configuration, as the diagonal transitions offer improved initialization times over the weakly allowed diagonal transitions in the Faraday configuration \cite{emary2007fast, lu2010direct}.

For each QD, the $\ket{\downarrow} \leftrightarrow \ket{\uparrow \downarrow \Downarrow}$ transition is coupled to the cavity mode, as required to perform both the two-qubit gates and readout of the electron spin states. The remaining transitions are far-detuned from the cavity, i.e. $\Delta E_e, \Delta E_g \gg \kappa$ where $\Delta E_e\,(\Delta E_g)$ is the excited (ground) states energy split and $\kappa$ is the cavity linewidth.

\begin{figure}[H]
	\centering
	\includegraphics[width=8cm]{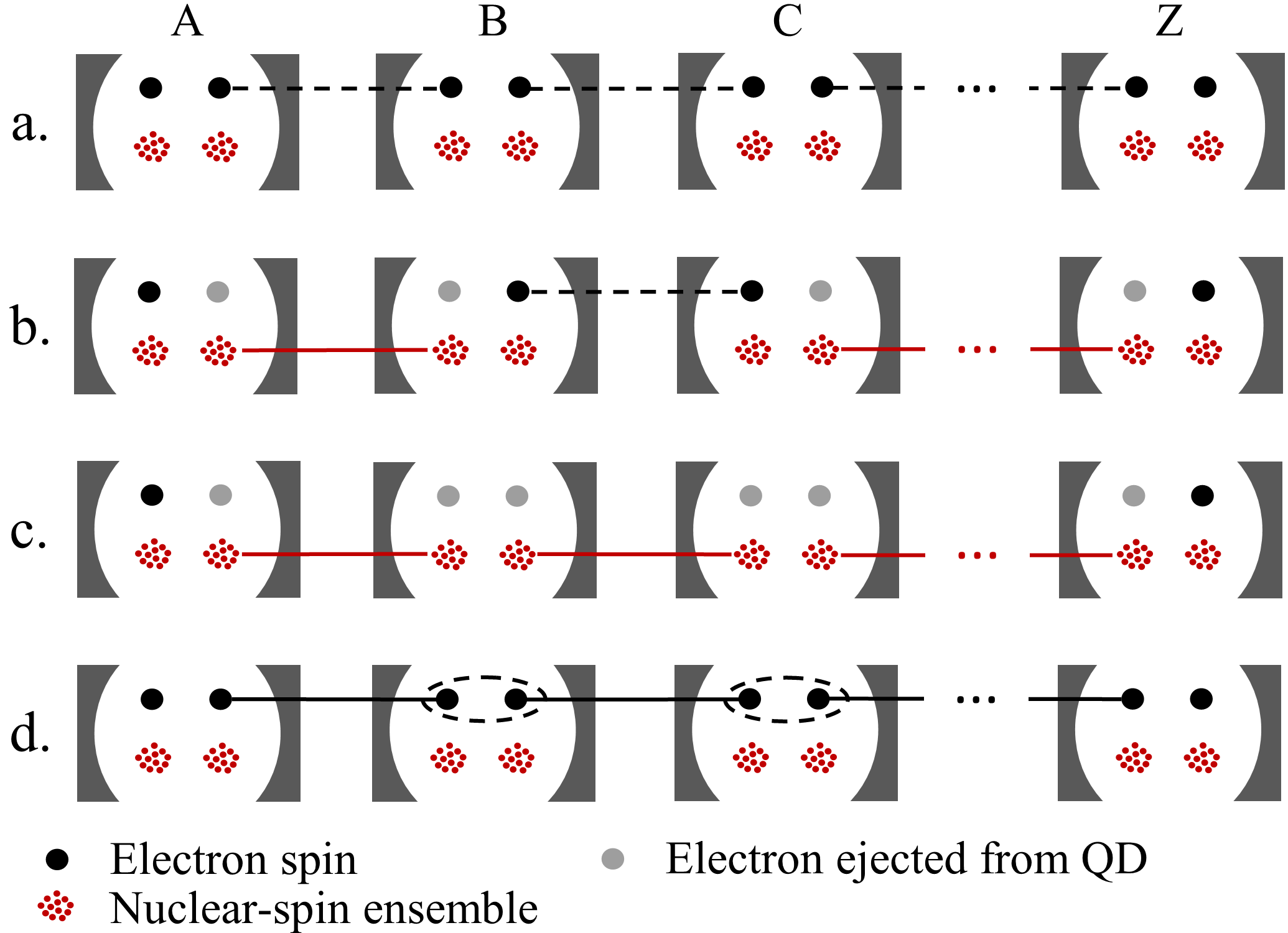}
	\caption{(a) To establish entanglement over a communication channel, entanglement generation is attempted between electron spins in neighboring nodes (dashed lines). Immediately following the entanglement generation pulses, each electron spin state is transferred to its corresponding nuclear-spin ensemble and the electron is ejected from the dot. (b) Photon detection heralds the establishment of entanglement between nuclear ensembles (solid lines), and entanglement generation is re-attempted where unsuccessful. (c) The result is an entangled state stored in the nuclear ensembles, in each of the local links. (d) The QDs are then recharged with electrons, the entangled state is transferred back to the electron spins, and entanglement is sequentially swapped using a cavity-enabled gate (dashed line), so as to extend the entanglement over the length of the neighboring local link. The electron spin and nuclear-spin ensemble are shown separately for illustration purposes. In reality the nuclear ensemble surrounds the electron spin.}
	\label{procedure}
\end{figure}

To distribute entanglement over the length of a communication channel, say between nodes $A$ and $Z$ depicted in Fig. \ref{procedure}, the first step of the protocol is to independently generate entanglement between QDs in nodes $A$ and $B$, $B$ and $C$, $\dots$, $X$ and $Y$, $Y$ and $Z$. Immediately following entanglement generation pulses, the electron spin state is transferred over to its nearby nuclear ensemble, and the electron is ejected from the QD, so as to extend the coherence time of the nuclei (Sec. \ref{initialization_sec}). Once entanglement is established in each of the local links, the QDs are recharged, the states of the nuclear ensembles are transferred back to the corresponding electrons, and entanglement is distributed over neighboring links by performing local gates between both QDs in an optical cavity.

\begin{figure}[H]
	\centering
	\includegraphics[width=8cm]{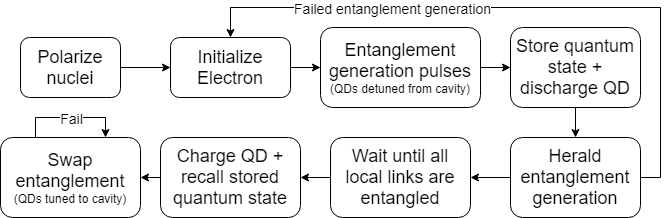}
	\caption{Flowchart representing the operations required for each node of the repeater protocol.}
	\label{flowchart}
\end{figure}

The repeater protocol can be described by the operations required in each local link, and the sequence of operations can be visualized using the flowchart shown in Fig. \ref{flowchart}. The following sections provide a discussion of each of the operations required by our repeater scheme.

\subsection{Initialization of electron and nuclear spins}
\label{initialization_sec}

The electron spin can be initialized to the $\ket{\downarrow}$ state through optical pumping of the $\ket{\uparrow} \rightarrow \ket{\uparrow\downarrow\Uparrow} $ transition. Spontaneous decay into the two ground states will initialize the electron spin to the $\ket{\downarrow}$ state within nanoseconds \cite{single_spins_QDs}. Once the spin is shelved in the $\ket{\downarrow}$ state it will not interact with the driving field due to the difference between the transition frequency and the laser frequency.

A requirement of the nuclear-spin-ensemble memory is for the ensemble to be initially polarized to perform the electron-nuclear state transfer with high fidelity. The polarization can be achieved via a process called dynamic nuclear spin polarization (DNSP) in which the electron acts as a probe to the nuclear spins. Spin angular momentum is transferred from the electron to the surrounding nuclear spins resulting in the polarization of the nuclei \cite{Nuclear_spins_QDs}. There has been a significant amount of research done in the past decade resulting in several methods to polarize QD nuclei \cite{Nuclear_spins_QDs}.

DNSP can be performed via nonresonant excitation with circularly polarized light. The work of Ref. \cite{chekhovich2017measurement} reported a nuclear polarization of $80\%$ in GaAs/AlGaAs QDs which is currently the largest achieved. Maximum polarization was attained through the use of a high powered, continuous-wave multimode laser tuned approximately $0.05$ eV above QD resonance. The experiment was performed in the Faraday configuration, however, various experiments have been performed using optical pumping with circularly polarized light in the Voigt configuration \cite{bracker2005optical, urbaszek2007efficient}. In light of the recent advances made towards high levels of nuclear polarization there is the potential that improved methods could lead the way to achieving nuclear polarizations close to $100\%$.

We will consider the case where DNSP was used to initialize the nuclear ensemble of $N$ spin-$I$ nuclei into the ideal state
\begin{equation}
\label{ideal_nuclear}
\ket{\bm{0}} = \ket{-I}_1 \otimes \ket{-I}_2 \otimes \dots \otimes \ket{-I}_N ,
\end{equation}
where $\ket{-I}_i$ denotes that the $i$-th nuclei ($i \in N$) is in its lowest energy state. The effect of partial initial nuclear polarizations will be discussed in terms of the electron-nuclear state transfer fidelity in Sec. \ref{fid_state_trans}. 

The nuclear memory coherence time is primarily determined by dephasing due to the electron-nuclear hyperfine interaction. The effect of this interaction can be either reduced by using electron-spin resonance spin-echo techniques, or eliminated by removing the electron from the dot after state transfer. Considering the latter, we estimate that dephasing of the nuclear spins due to the hyperfine interaction is negligible over a nanosecond timescale. As such, the nuclear spins are expected to remain coherent if the electron can be added/removed from the dot over such a timescale. Progress has been made in this direction, and it has been demonstrated that the charge state of an InGaAs QD can be manipulated in less than 1.4 ns \cite{nannen2010ultrafast}.

When the QD is uncharged, the dominant decoherence process is spin diffusion via the nuclear dipole-dipole interaction. On one hand, it is known that quadrupolar shifts resulting from inhomogeneous strain will reduce the rate of nuclear spin diffusion, which is particularly true for highly-strained InGaAs QDs \cite{chekhovich2015suppression}. On the other hand, a high-fidelity realization of the state transfer process requires a narrow distribution of inhomogeneous quadrupolar shifts (see Sec. \ref{Subsection_mapping}). To that end, Ref. \cite{ulhaq2016vanishing} has shown that while the levels of strain in GaAs/AlGaAs NDFE QDs are several orders of magnitude smaller than those of InGaAs QDs, there is indeed sufficient variance in the strain to obtain comparable diffusion rates.

NMR techniques can also be used to mitigate the effects of nuclear spin diffusion. These techniques are well-developed and can lead to nuclear $T_2$ approaching the second timescale \cite{Long_lived_memory, controlling_mesoscopic_spin_envir}. It is therefore reasonable to consider nuclear coherence times on the second timescale for an uncharged dot, which are sufficient to complete the repeater protocol for typical lengths of quantum repeaters ($500$ km - $2000$ km).

\subsection{Entanglement generation}

The generation of entanglement between the two QD electrons in a local link of the repeater can be accomplished using the scheme proposed by Barrett and Kok \cite{Barrett_Kok} which has been used in practice to generate entanglement between two electron spins in diamond, separated by 1.3 km \cite{hensen2015loophole}.

First, the spin state of the electron in each dot is initialized to the $\ket{\downarrow}$ state through the use of optical pumping, as in the previous section. The electron spin can then be coherently rotated to the superposition state $\frac{1}{\sqrt{2}} ( \ket{\uparrow} + \ket{\downarrow} )$ using ultrafast pulses of light \cite{press2008complete}. The application of a laser pulse which is resonant with the $\ket{\downarrow} \rightarrow \ket{\uparrow\downarrow\Downarrow}$ transition, followed by spontaneous emission, will entangle each electron spin state with the emitted photon number. 

The Purcell-enhanced lifetime of the cavity-coupled transition is expected to be shorter than the excitation pulse width required by the time-bandwidth product to resolve the energy level splitting, which can lead to the emission of multiple photons for a single excitation pulse. In order to resolve the energy levels and prevent re-excitation, we propose that the QDs are detuned from cavity resonance to reduce the effective Purcell enhancement.

For QDs with an optical lifetime on the order of a nanosecond in a magnetic field of several Telsa, the effective Purcell enhancement will be required to be around 1-18 (discussed in detail in Sec. \ref{implementation_entanglement_generation}). Purcell factors in excess of 100 are required for a high-performance implementation of the two-qubit gates, and as such, the QD should only be detuned from the cavity resonance during the entanglement generation step. Since the $\ket{\uparrow \downarrow \Downarrow}$ state can decay to either of the electron ground states, we filter out the $H$-polarized photons at the expense of efficiency. Thus, when the electron is in the state $\ket{\downarrow}$ ($\ket{\uparrow}$) there will be $1$ ($0$) emitted photon(s).

The photons are then directed towards a beam splitter (BS) which is located at a mid-point between the two nodes. The detection of a single photon in one of the two BS output modes will project the electron spins onto an entangled state. It is highly likely that photon loss will occur when the spontaneously-emitted photons are directed towards the BS. This can lead to the situation where both QDs in the local link emit a photon, yet only one photon is detected. In this situation, the electrons will be left in a product state rather than an entangled state. To eliminate such a possibility, the electron spins are flipped after the first excitation-emission and a second excitation pulse is applied.

The detection of two consecutive single photons at the BS will leave the electron spins of dots $D_{i}$ and $D_{i+1}$ in the maximally entangled state given by $ \ket{\Psi^\pm} = \frac{1}{\sqrt{2}} ( \ket{\uparrow_{i} \downarrow_{i+1}} \pm \ket{\downarrow_{i} \uparrow_{i+1}} )$, where the $+(-)$ sign corresponds to the case where the same (different) detector(s) received a photon.

As mentioned in Sec. \ref{initialization_sec}, we require the nuclear spin coherence times offered by uncharged QDs. To that end, immediately following the second excitation pulse, the electron spin state is transferred to its nearby nuclear-spin ensemble and the electron is ejected from the dot. As a result of this additional state transfer step in our protocol, the detection of two photons will project the nuclear ensembles onto an entangled state, rather than the electron spins.

The entanglement generation process is performed in each of the local links of the repeater such that the final result is the generation of piece-wise entanglement across the length of the channel.

\subsection{State Transfer}
\label{Subsection_mapping}

The spin state of an electron confined in a QD can be transferred to the surrounding nuclear spins of the dot through the strain-induced, noncollinear hyperfine interaction \cite{non_ideal_polarization}. State transfer with high fidelity requires that the variance in inhomogeneous quadrupolar shifts be less than $100$ kHz (Sec. \ref{fid_state_trans}), which is attainable for low-strain GaAs/AlGaAs QDs \cite{chekhovich2017measurement}. For reference, the quadrupolar shifts in InAs QDs are strongly inhomogeneous and are on the order of several MHz \cite{chekhovich2012structural, bulutay2012quadrupolar}. As such, we propose that low-strain GaAs/AlGaAs QDs are nearly homogeneously strained using a piezoelectric actuator \cite{trotta2015energy, trotta2016wavelength}.

Transferring the electron spin state to the nuclear ensemble will extend the coherence time of the quantum state, providing the time required for entanglement to be generated in all of the local links. The transfer of a single unit of angular momentum to the nuclear ensemble corresponds to the creation of a single nuclear magnon, the elementary quantum unit of a collective nuclear-spin wave. We will consider the nuclear-spin-wave operators $\Phi_{\Delta m}^\pm$, $\Delta m \in \{1,2\}$, which change the net nuclear spin by $\pm \Delta m$.

A nuclear spin transition corresponding to the nuclear-spin-wave operator can be switched on by a Hamiltonian-engineering pulse sequence on the electron spin. Following the method used in Ref. \cite{non_ideal_polarization}, which is described in detail in Ref. \cite{mapping_pulse_seq}, the electron spin is driven with a series of short $S_x$ and $S_y$ pulses separated by a time interval $\tau$. The spin rotations can be carried out using an all-optical Raman drive. The coupling between the electron and the $\Delta m$-mode is resonantly enhanced by setting $\tau = 3 \pi / (4\omega_Z^n \Delta m)$, where $\omega_Z^n$ denotes the nuclear Zeeman frequency splitting, resulting in a targeted change in nuclear spin of $\Delta m$.

For nuclear spin species with sufficiently different Zeeman energies, as is the case for GaAs QDs, the pulse sequence can selectively transfer the electron spin state to the targeted nuclear species. The system will then evolve under an effective flip-flop Hamiltonian
\begin{equation}\label{Hamiltonian}
H = A_{\Delta m}^\prime ( \Phi_{\Delta m}^{+}S_{-} + \Phi_{\Delta m}^{-}S_{+} ) ,
\end{equation}

\noindent where $S_{\pm} = S_x \pm \text{i} S_y $ and $A_{\Delta m}^\prime$ is a re-scaled hyperfine coupling rate. The pulse sequence provides the ability to use the nuclear-spin ensemble as the quantum memory that is required in our repeater scheme. If we assume that the nuclear ensemble was initially polarized to the ideal state $\ket{\bm{0}}$, and that the electron is in an arbitrary state $| e \rangle = \alpha \ket{\uparrow} + \beta \ket{\downarrow} $, then the system will evolve as

\begin{equation}
\begin{split}
\ket{\psi(t)} = \alpha [ \cos (g_{\Delta m}t) \ket{\uparrow} &\otimes \ket{\boldsymbol{0}} \\
- \text{i} \sin (g_{\Delta m} t) \ket{\downarrow} &\otimes \ket{\boldsymbol{1}} ] + \beta \ket{\downarrow} \otimes \ket{\boldsymbol{0}} ,
\end{split}
\end{equation}

\noindent where $g_{\Delta m} $ is a collectively enhanced noncollinear coupling rate \cite{non_ideal_polarization}, and $\ket{\bm{1}} \propto \Phi_{\Delta m}^{+} \ket{\bm{0}}$. The state of the electron-nuclear system after a time $t=\pi/(2g_{\Delta m})$ is then

\begin{equation}
\ket{\psi(\pi/2g_{\Delta m})} = \ket{\downarrow} \otimes (-\text{i} \alpha \ket{\boldsymbol{1}} + \beta \ket{\boldsymbol{0}} ) ,
\end{equation}

\noindent and the state of the electron spin has been transferred to the nuclear-spin ensemble. Using this scheme, for example, it is possible for the state of entangled electron spins in dots $D_i$ and $D_{i+1}$ to be transferred over their nuclear-spin ensembles as
\begin{equation}
\begin{aligned}
\frac{1}{\sqrt{2}} ( \ket{\uparrow_i \downarrow_{i+1}} \pm \ket{\downarrow_i \uparrow_{i+1}} ) &\ket{\boldsymbol{0}_i\boldsymbol{0}_{i+1}} \\
\rightarrow -\text{i} \ket{\downarrow_i \downarrow_{i+1} }\frac{1}{\sqrt{2}} (&\ket{\boldsymbol{1}_i\boldsymbol{0}_{i+1}}\pm\ket{\boldsymbol{0}_i\boldsymbol{1}_{i+1}}) .
\end{aligned}
\end{equation}

If both electron spins are in the $\ket{\downarrow}$ state, the evolution of the system permits the original state of the electron spin to be transferred back to the electron from the nuclear ensemble using the same sequence of pulses. Thus, once entanglement has been generated in each of the local links, the entangled nuclear states can be transferred to the electron spins and the system is ready for the entanglement swapping process.

\subsection{Entanglement swapping}
\label{entanglement_swapping}

We can distribute entanglement across neighboring links of the repeater through the use of the cavity-assisted photon scattering (CAPS) scheme described in Ref. \cite{photon_scattering_gate}. This scheme has been realized experimentally using rubidium atoms separated by several micrometers inside an optical cavity \cite{welte2018photon}. A controlled phase-flip gate between electron-spin qubits in the same cavity can be performed by scattering a single photon off of the cavity and detecting it. Measurement of the reflected photon both heralds the gate and makes the fidelity of the gate robust against photon loss.

We propose the use of CAPS-based gates because the QDs can be separated by several hundred nanometers \cite{calic2017deterministic}, whereas performing the gate via the electric dipolar interaction \cite{Simon_qds} requires that the QDs be situated relatively close together, around $20$ nm. Additionally, experimental demonstrations have shown that QDs are promising on-demand single-photon sources with high single-photon purity and indistinguishability \cite{unsleber2016highly, liu2018high, somaschi2016near}. Implementation of these gates through the use of other QD-cavity single-photon sources represents a promising route towards scalable, on-chip quantum logic \cite{qd_phc_reflec1, qd_phc_reflec2, cao2019high, wei2014scalable}. 

A requirement for the photon scattering gate is a high-cooperativity cavity, $C = 4g^2/\kappa\gamma \gg 1$ where $\kappa$ is the cavity decay rate, $g$ is the cavity coupling rate, and $\gamma$ is the decay rate of the quantum system excited state. Note that here emitter pure dephasing is neglected. High cooperativity is achievable in both the bad-cavity regime where $\kappa \gg g \gg \gamma$ and the strong-coupling regime where $g \gg \kappa \gg \gamma$ \cite{Faezeh-gate}. 

This scheme also requires a single-sided cavity. For both QD-electron systems, the $\ket{\downarrow} \leftrightarrow \ket{\uparrow\downarrow\Downarrow}$ transition is resonant with the cavity and the $\ket{\uparrow} \leftrightarrow \ket{\uparrow\downarrow\Uparrow}$ transition should not interact with the cavity, as shown in Fig \ref{e_levels}. Since both ground states can interact with the excited state with linearly polarized light, the $\ket{\uparrow} \leftrightarrow \ket{\uparrow\downarrow\Uparrow}$ transition should be far detuned from the cavity frequency.

When the frequency of the photon is resonant with the bare cavity mode and if both electrons are in the spin state $\ket{\uparrow}$, an incident photon would enter and reflect back from within the cavity. This reflection causes the state of the electron-photon system to acquire a $\pi$-phase shift.
Alternatively, when either or both of the electrons are in the spin state $\ket{\downarrow}$, the reflection properties of the cavity are modified \cite{scattering_reflection_prop}. In this case, the cavity acts as a mirror and the photon is reflected back before entering the cavity. Under the ideal conditions, the system acquires no phase change.

The result of the photon scattering scheme is that there is a phase flip in the system only if both electrons are in the state $\ket{\uparrow}$. The cavity-assisted photon scattering scheme realizes a controlled-Z (CZ) gate \cite{nielsen_chang}:
\begin{equation}
\begin{split}
& \ket{\downarrow\downarrow} \rightarrow \ket{\downarrow\downarrow} \qquad \ket{\downarrow\uparrow} \rightarrow \ket{\downarrow\uparrow} \\
& \ket{\uparrow\downarrow} \rightarrow \ket{\uparrow\downarrow} \qquad \ket{\uparrow\uparrow} \rightarrow -\ket{\uparrow\uparrow} .
\end{split}
\end{equation}

Any two neighboring links in the repeater consist of 4 QDs, which can be denoted by $D_1, D_2, D_3$ and $D_4$. Entanglement is generated between electron spins in dots $D_1$ and $D_2$, as well as between dots $D_3$ and $D_4$.

For demonstration purposes, we assume that the entangled states are both given by $\ket{\Psi^+}$. To perform the entanglement swapping between nearby dots, i.e. $D_2$ and $D_3$, one needs to perform a controlled NOT (CNOT) gate between them followed by state measurement of the electron spins. A CNOT gate can be performed by first applying a Hadamard gate on $D_2$. Then, a CZ gate, as explained before, changes the phase of the joint state of $D_2$ and $D_3$. Finally, another Hadamard gate is performed on $D_2$. After performing the CNOT gate, we measure $D_2$ in the Z basis and $D_3$ in the X basis. The latter is obtained by applying a Hadamard gate on $D_3$ and then measuring it in Z basis.

Depending on the measurement results, the entangled state between $D_1$ and $D_4$ will be projected onto one of the Bell states. Entanglement swapping can be performed simultaneously with other links of the repeater. The swapping procedure is repeated until an entangled state has been generated between the electron spins which are located at the endpoints of the transmission channel.

\subsection{State Measurement}

Measurement of the electron-spin state is performed by exciting the system in a spin-selective manner and detecting the fluorescence. We propose the use of the AC Stark effect to adjust the energy levels of the QD and obtain a cycling transition \cite{AC_Stark_proposal, wilkinson2019dynamic}. The application of a strong, far-detuned, circularly polarized laser to a QD in the Voigt configuration results in an energy level structure similar to the Faraday configuration. In this pseudo-Faraday configuration, there are two strong spin-preserving transitions and two weak spin-flipping transitions. The two spin-preserving transitions are nondegenerate, hence, resonant excitation of one of them is a spin-selective excitation. A projective measurement of the QD electron spin may be accomplished by a laser tuned into resonance with the cavity-coupled $\ket{\downarrow} \leftrightarrow \ket{\uparrow \downarrow \Downarrow}$ transition. It is therefore necessary to tune the $\ket{\downarrow} \leftrightarrow \ket{\uparrow \downarrow \Downarrow}$ transition into resonance with the cavity after application of the AC Stark laser.

If the laser is used to probe the QD from above the QD-cavity plane, then the emission will primarily be into the cavity mode which is in-plane and orthogonal to the probe laser direction. Purcell-enhanced emission is possible in this geometry as demonstrated by an InGaAs QD coupled to a photonic crystal cavity \cite{liu2018high}. Detection of one or more photons leaves the qubit in the $\ket{\downarrow}$ state, otherwise it is left in the $\ket{\uparrow}$ state.

\section{Entanglement generation rates}
\label{Rates}

The performance of a quantum repeater scheme can be quantified both by the rate at which entanglement is distributed over the length of the communication channel and by the fidelity of the generated entanglement. The entanglement generation rate of our scheme is discussed in this section.

In a magnetic field of several Tesla, it is expected that the nuclear polarization build-up time is $1-10$ s \cite{chekhovich2013nuclear}, the nuclear spin lifetime is several hundred seconds \cite{ulhaq2016vanishing} and the coherence time is on the order of seconds when the dot is uncharged. If the polarization decay processes as a function of time is described by a simple exponential \cite{maletinsky2007dynamics}, and the polarization decay rate is comparable to the nuclear spin lifetime, then polarization decay is negligible over the coherence time of the nuclei. We consider the case where DNSP is performed at the start of the protocol, prior to the first entanglement generation attempt in each local link.

High-quality QD telecom sources are not yet available so our scheme will likely require the conversion of $\sim 750$ nm photons emitted from the GaAs QDs to the telecom wavelength range (1.3-1.5 $\mu$m) to match the low-loss wavelength range of silica fibers. We discuss frequency conversion in detail in Sec. \ref{freq_con} and assume here that the conversion efficiency is ideal.

The success probability of generating an entangled state between two QD electrons, separated by $L_0$, is $p_0 = \frac{1}{2} (\zeta \eta_t p \eta_c \eta_d)^2$, where $\zeta$ is the Purcell-enhanced branching ratio \cite{casabone2021dynamic} in the Voigt geometry, $\eta_t = e^{-L_0/(2 L_{att})}$ is the transmission probability of a photon subject to fiber attenuation of $0.17$ dB/km ($L_{att} \approx 25$ km), $p$ is the probability of emitting a photon into the cavity mode, $\eta_c$ is the collection efficiency, and $\eta_d$ is the detection efficiency \cite{atomic_ensembles}.

In our protocol, immediately after the second entanglement generation optical $\pi$ pulse, the spin state of each electron is transferred to the surrounding nuclear ensemble and the electron is ejected from the dot to eliminate electron-mediated decoherence of the nuclear spins. If the entanglement generation attempt is unsuccessful, then the QD must be recharged and the nuclear and electron spins must be reinitialized. Reinitialization of the nuclei is essential to ensure that the nuclear polarization is maintained throughout the entanglement generation process. Manipulating the charge state of a QD requires $\sim 1$ ns \cite{nannen2010ultrafast} and the electron spin can be reinitialized within several nanoseconds (Sec. \ref{initialization_sec}). The change in the nuclear polarization due to the electron state transfer can be reversed by reinitializing both electrons to the $\ket{\downarrow}$ state and performing a second application of the state transfer pulse sequence, such that the collective nuclear spin excitation is transferred back to the corresponding electron. Both electrons are then reinitialized and rotated into the superposition state, and the entanglement generation step is again attempted. For a nuclear polarization of $95\%$ the state transfer time is at most $165$ ns \cite{non_ideal_polarization}  using the parameters discussed in Sec. \ref{fid_state_trans}. The total time required to reinitialize the electron-nuclear system after a failed entanglement generation attempt is estimated to be $\tau_\text{init} \approx 0.2\,\mu$s. The average time of generating and storing entanglement in a local link is
\begin{equation}
\langle T \rangle_{L_0} = \Bigg( \frac{L_0}{c} + \tau_\text{init} \Bigg) \frac{1}{p_0} ,
\end{equation}
\noindent where $c = 2 \times 10^8$ m/s is the speed of light in an optical fiber. An approximation for the time required for entanglement generation in two neighboring links is given by $3/2 \langle T \rangle_{L_0}$ \cite{atomic_ensembles}.

Just after generating entanglement in two neighboring elementary links, we recharge the dots, transfer the collective nuclear state back to the corresponding electron, perform entanglement swapping, and store the entangled state in the nuclear ensembles. These times are quite negligible compared to the time required to establish two neighboring links.

After completing the CNOT gate, the states of the two electron spins are measured. These measurements are deterministic and the time required to perform them is negligible (Sec. \ref{Section_readout}). The success probability for the entanglement swap is then $p_s \approx p_\text{gate}$, where $p_\text{gate}$ is the success probability of performing the photon scattering gate. Factors which limit $p_\text{gate}$ include the efficiencies of single-photon sources, single-photon detectors, and optical losses of the circuit. An approximation for the gate probability is given by $p_\text{gate} = \eta_\text{s} \eta_c \eta_\text{cav} \eta_d$ where $\eta_\text{s}$ is the source efficiency and $\eta_\text{cav}$ includes the inefficiencies in the cavity.

The average time to distribute entanglement over two neighboring links is
\begin{equation}
\langle T \rangle_{2L_0} = \frac{3}{2} \frac{L_0/c +\tau_\text{init}}{p_0 p_s} .
\end{equation}

If the two QDs in each cavity can be individually addressed or filtered by wavelength such that emission is directed to the associated BS, then entanglement generation attempts can be performed on neighboring links at the same time. For a communication channel with length $L$ and $l=L/L_0$ local links, entanglement can be created over the distance $L$ with $n$ levels of entanglement swapping operations. We refer to $n$ as the nesting level. The average time to distribute an entangled pair over a distance $L = 2^n L_0$ is then
\begin{equation}
\label{time}
\langle T \rangle_L = \Big( \frac{3}{2} \Big)^n \frac{L_0/c + \tau_\text{init}}{p_0 p_s^n} .
\end{equation}
If one needs to generate entanglement over neighboring links one-by-one, the above expression changes to
\begin{equation}
\langle T \rangle_L = 2 \, \Big( \frac{3}{2} \Big)^{n-1}\,  \frac{L_0/c + \tau_\text{init}}{p_0 p_s^n}.
\end{equation}

We chose to investigate the dependence of the entanglement distribution rate on extraction efficiency as there has been significant progress in recent years made towards the collection of single-photon emission from QDs \cite{dalacu2019nanowire, arakawa2020progress}. We compare the rates associated with several implementations of our scheme to a repeater scheme based on a deterministic photon-pair source and two-photon Bell-state measurement (referred to as a $2+2$ scheme \cite{wu2020near}). Such a scheme is particularly relevant for GaAs QDs which can generate photon pairs via the biexciton-exciton cascade with low multi-photon probability and high degrees of entanglement and indistinguishability. Alternatively, a comparison could have been made to a scheme based on a single-photon source with a single-photon Bell-state measurement (a $1+1$ scheme \cite{wu2020near}). Such a scheme is relevant for QD-based repeaters which utilize external quantum memories. We focus on the $2+2$ scheme as the corresponding entanglement generation rates are expected to exceed those of a $1+1$ scheme \cite{wu2020near}.

The average time to distribute entanglement using a $2+2$ scheme is given by \cite{wu2020near}
\begin{equation}
\label{2+2}
\langle T \rangle_L^\prime = \Big( \frac{3}{2} \Big)^n \frac{L_0 / c}{p_0^\prime {p_s^\prime}^n} ,
\end{equation}
where $p_0^\prime = 0.5 (\eta_t \eta_s \eta_d )^2$ is the probability of generating entanglement in a single link, $p_s^\prime = 0.5 \eta_d^2 \eta_m^4$ is the entanglement swapping probability, and $\eta_m$ is the memory efficiency. Eqn. \ref{2+2} is valid when memory initialization times and memory decay are negligible over the timescales involved in the repeater protocol.

\begin{figure}
	\centering
	\includegraphics[trim=0cm 0cm 1cm 1cm, clip=true, width=8.6cm]{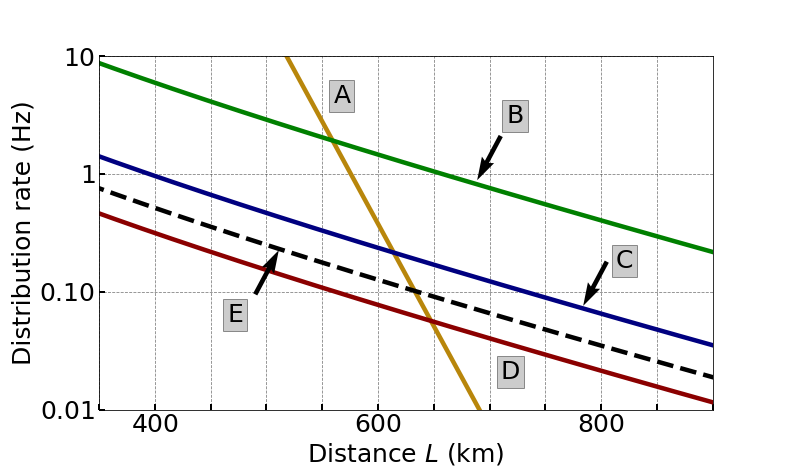}
	\caption{Comparison of entanglement generation rates as a function of distance for (A) direct transmission using a $10$ GHz single-photon source, and our QD-based repeater scheme for (B) $p\eta_c = 0.72$, (C) $p\eta_c = 0.5$, (D) $p\eta_c = 0.4$. We considered $\eta_d=\eta_\text{cav}=0.9$, a nesting level of $n=3$, and an effective Purcell factor of $F_p = 16$ during the entanglement generation step. We also compare to (E) a repeater scheme based on deterministic photon-pair source and two-photon Bell state measurement, with source efficiency $\eta_s = 0.65$ and memory efficiency $\eta_m = 0.9$.}
	\label{fig:repeater_rates}
\end{figure}

The entanglement distribution rates of our scheme (Eqn. \ref{time}) are plotted in Fig. \ref{fig:repeater_rates} as a function of distance for various values of $p\eta_c$, and are compared to direct transmission (curve A) using a single-photon source which produces photons at 10 GHz. We also compare to a $2+2$ scheme (curve E) with a photon-pair source efficiency $\eta_s = 0.65$ \cite{liu2019solid} and high memory efficiency $\eta_m = 0.9$, possible, for example, using a gradient echo memory scheme \cite{hosseini2011high}. A high-performance implementation of our scheme (curve B) can be envisioned by considering an extraction efficiency of $p \eta_c = 0.72$, measured for a single InAs QD embedded in a GaAs photonic nanowire \cite{claudon2010highly}, including the probability of emitting into the nanowire and the nanowire-detector coupling. We have also plotted $p \eta_c = 0.5$ (curve C) and $p \eta_c = 0.4$ (curve D). The source efficiency, when implementing the gate using a QD single-photon source, can be approximated as $\eta_s = p$. Here we consider $\eta_\text{cav} = 0.9$ \cite{faraon2007efficient}, $\eta_d = 0.9$, and a branching ratio of $\zeta= 0.94$ corresponding to an effective Purcell enhancement of $F_p=16$ (Sec. \ref{entanglement_gen_fid}). For $p \eta_c = 0.72$, $p \eta_c = 0.5$, and $p \eta_c = 0.4$, we calculate $p_\text{gate} = 0.58$, $p_\text{gate} = 0.41$, $p_\text{gate} = 0.32$, respectively.
	
Fig. \ref{fig:repeater_rates} shows that extraction efficiencies of $p \eta_c \gtrsim 0.5 $ are required if our repeater scheme is to outperform the $2+2$ scheme. While such efficiencies are currently out-of-reach for GaAs/AlGaAs QDs, they are achievable for self-assembled QDs \cite{claudon2010highly, liu2017deterministic}. The work being done on entangled photon-pairs from GaAs QDs will naturally lead to the desire to incorporate these dots in microcavities. As such, we expect rapid development in nanofabrication techniques resulting in QD-cavity metrics for low-strain GaAs QDs that will make it possible to realize the high-performance implementation of our scheme presented above. We further discuss this in Sec. \ref{Implementation}. Finally, it is expected that adapting our scheme to incorporate multiplexing \cite{collins2007multiplexed, asadi2018quantum, atomic_ensembles} would increase the achievable entanglement generation rates.

\section{Fidelity}
\label{Fidelity}
In this section we present the fidelity associated with each of the components of our repeater protocol as well as the overall fidelity. The computations presented in the following sections consider the fidelity as $F=\braket{\psi | \hat{\rho} | \psi}$, where $\ket{\psi}$ is the desired pure state and $\hat{\rho}$ is the evolved state.

\subsection{Entanglement generation}
\label{entanglement_gen_fid}

To calculate the fidelity of the entanglement generation step, we assume that the detection window is much longer than the optical decay time of the system and that the spin decoherence is negligible over the timescale of the lifetime of the QD. Our scheme requires that the electron spin coherence time can be extended from several hundred nanoseconds to several microseconds. We discuss various methods of satisfying this requirement in Sec. \ref{implementation_entanglement_generation}. The fidelity of the Barrett and Kok entanglement generation scheme is then given by \cite{wein2020analyzing, asadi2020protocols}:
\begin{equation}
\label{barrett-kok}
F_\text{BK}= \frac{1}{2}\left[ 1 + \frac{4 \gamma_i^{\prime} \gamma_{i+1}^{\prime} }{(\Gamma_i^\prime + \Gamma_{i+1}^\prime)^2 + 4 \delta_\omega^2}\right],
\end{equation}
where $\gamma_k^\prime = \gamma_{\text{r},k} (1+F_{\text{p},k})+\gamma_{\text{nr},k}$ is the cavity-enhanced optical decay rate of the $k$th QD ($k\in\{i,i+1\}$), $F_{\text{p},k}$ is the effective Purcell factor, $\gamma_{\text{r},k}$ ($\gamma_{\text{nr},k}$) is the radiative (non-radiative) decay rate, $\Gamma_k^\prime=\gamma_k^\prime + 2 \gamma_k^\star$ is the spectral full width at half maximum (FWHM) of the homogeneously-broadened zero-phonon line, $\gamma_k^\star$ is the optical pure dephasing and $\delta_\omega$ is the detuning between the optical transition frequencies $\omega_i$ and $\omega_{i+1}$. Eqn. \ref{barrett-kok} is valid when the infidelity due to re-excitation emission and phonon sideband emission is negligible compared to that of pure dephasing and spectral diffusion.

To resolve the $\ket{\downarrow} \leftrightarrow \ket{\uparrow \downarrow \Downarrow}$ transition and prevent re-excitation, the QDs should be detuned from cavity resonance during entanglement generation. The Purcell factor as a function of the detuning between the $k$th QD's transition frequency, $\omega_k$, and the cavity frequency, $\omega_\text{cav}$, can be expressed as
\begin{equation}
    F_p = \frac{\kappa^2}{4(\omega_k - \omega_\text{cav})^2 +\kappa^2} F_\text{res} \, ,
\end{equation}
where $\kappa$ is the cavity linewidth and $F_\text{res}$ is the Purcell factor when the QD and cavity are resonant \cite{casabone2021dynamic}. For $F_\text{res} = 500$ ($F_\text{res} = 200$), detuning the QD transition frequency from cavity resonance by $2\pi \times 275$ GHz ($2\pi \times 200$ GHz) \cite{hohenester2009phonon, laucht2009electrical} will result in $F_p = 16 $ ($F_p = 12$) for a cavity decay rate of $\kappa = 2\pi \times 100$ GHz \cite{liu2018high} (the cited detunings and cavity decay rate are for InGaAs QDs coupled to photonic crystal cavities).

If we consider GaAs/AlGaAs QD samples with negligible nonradiative decay channels ($\gamma_{nr} \ll \gamma_r$), then $\gamma \approx \gamma_{r} = 2\pi \times 0.59$ GHz \cite{zhai2020low}. We then use $\Gamma = 0.64$ GHz \cite{zhai2020low} to calculate the cavity-enhanced optical decoherence rate for a specific Purcell factor.

To account for spectral diffusion we follow Ref. \cite{wein2020analyzing} and average Eqn. \ref{barrett-kok} over a Gaussian distribution $f(\omega_k - \overline{\omega}_k, \sigma_k) = (\sigma_k \sqrt{2\pi})^{-1} e^{-(\omega_k - \overline{\omega}_k)^2/2 \sigma_k^2} $ for each QD frequency $\omega_k$ with an average value of $\overline{\omega}_k$ and a spectral diffusion standard deviation $\sigma_k$. The entanglement generation fidelity between QDs $D_i$ and $D_{i+1}$ is then given by 
\begin{equation}
\begin{aligned}
F_\text{ent} = \iint &f(\omega_i - \overline{\omega}_i, \sigma_i) \\ \times &f(\omega_{i+1} - \overline{\omega}_{i+1}, \sigma_{i+1}) F_\text{BK}\, d\omega_i d\omega_{i+1} .
\end{aligned}
\end{equation}

For a Purcell factor of $F_p=500$ ($F_p = 200$), $\kappa = 2\pi \times 100$ GHz, $\overline{\omega}_i = \overline{\omega}_{i+1} = \omega_\text{cav} + 2\pi \times 275$ GHz ($200$ GHz) and spectral diffusion FWHM of $2\pi \times 500$ MHz (discussed in Sec. \ref{Implementation}) for both QDs, the fidelity of the entanglement generation step is $F_\text{ent}=0.995$ ($F_\text{ent} = 0.993$).

\subsection{State Transfer}
\label{fid_state_trans}

 A numerically exact technique for calculating the dynamics of the electron-nuclear state transfer and the corresponding fidelity was developed by Denning \textit{et al.} in Ref. \cite{non_ideal_polarization}. The spin state of the nuclear ensemble is mapped onto two one-dimensional chains of states. Each of these chains represent the set of states corresponding to the evolution of either the positive or the negative spin wave. A transition from one state in the chain to a neighboring state represents a spin flip-flop between the electron and the nuclear ensemble. The one-dimensional structure of these chains can be attributed to the secular form of the interaction Hamiltonian given by Eqn. \ref{Hamiltonian}, which results in non-zero coupling between only neighboring nuclear states in a chain.

The calculated fidelity considers a full write-read cycle and is averaged over six initial electron states $ \alpha \ket{\uparrow} + \beta \ket{\downarrow}$, where $(\alpha, \beta) = (1, 0), \, (0, 1), \, \frac{1}{\sqrt{2}}(1, \pm 1), \,\frac{1}{\sqrt{2}}(1, \pm i)$.  In the following, we consider the fidelity calculations to include the effects of non-ideal electron spin initialization, partial nuclear polarization, and quadrupolar inhomogeneity. We assume that decoherence of the electron spin is negligible over the timescale of the state transfer and further discuss this matter in Sec. \ref{implementation_entanglement_generation}.

For a nuclear polarization of $80\%$ and an external magnetic field of $B_x = 6.6$ T, the ground and excited states are split by $\Delta E_g = 32$ GHz  and $\Delta E_e = 146$ GHz, respectively (Sec. \ref{implementation_entanglement_generation}). The relatively large splittings reduce possible off-resonant excitation when the electron spin is initialized via optical pumping. As a result, the electron spin can be initialized to the $\ket{\downarrow}$ state with a fidelity of $F_\text{e,init} = 0.99996$, following Ref. \cite{emary2007fast} for $\gamma_r = 2\pi \times 0.59$ GHz \cite{zhai2020low}.

In the case of partial nuclear polarization, the asymmetry of the ideal and non-ideal coupling rates, corresponding to the ideal and non-ideal spin wave evolution, allows for a high-fidelity realization of this quantum memory at realistic mean nuclear polarizations. The fidelity associated with the $\Delta m=2$ mode exceeds that of the $\Delta m=1$ mode due to differences in the coupling strength to non-ideal states. In the following analysis we will consider the $\Delta m = 2$ mode, which can be obtained using the pulse sequence described in Sec. \ref{Subsection_mapping}.

We extract the effect of partial nuclear polarization on the state transfer fidelity, by reproducing the electron-nuclear dynamics and fidelity calculations described by Denning \textit{et al} \cite{non_ideal_polarization}. For a full write-read cycle, a nuclear polarization of $95\%$ ($80\%$) will contribute a multiplicative factor of $F_\text{n,init}=0.998$ ($F_\text{n,init}=0.977$) to the state transfer fidelity, considering arsenide nuclei with gyromagnetic ratios of $g = 2\pi \times 7.22$ MHz T$^{-1}$ \cite{non_ideal_polarization}, and a magnetic field of $B_x = 6.6$ T.

The presence of quadrupolar inhomogeneities can lead to the build-up of a relative phase among the individual nuclear spin components, rotating the $\ket{\boldsymbol{1}}$ nuclear state into a set of excitations which do not interact with the electron. The individual quadrupole energy shifts are given by $\Delta_Q^i = B_Q^i (\cos^2 \theta^i -\frac{1}{2} \sin^2 \theta^i )$ \cite{non_ideal_polarization}, where $B_Q^i$ is the quadrupolar interaction strength of the $i$-th nucleus ($i\in N$) and $\theta^i$ is the angle between the $i$-th nuclear quadrupolar axis and the direction of the applied magnetic field.

If we assume that the quadrupolar shifts are normally distributed with standard deviation $\sigma(\Delta_Q^i)$, then the population of the nuclear state $\ket{\boldsymbol{1}}$ will decay according to $\exp[ - \Delta m^4 \sigma(\Delta_Q^j)^2 t_\text{w/r}^2 ] $ \cite{non_ideal_polarization}, where $t_\text{w/r}$ is the timescale of a write-read cycle. Decoherence of the $\ket{\boldsymbol{1}}$ state due to quadrupolar inhomogeneities is only an issue during state transfer, as it has been suggested that the nuclear spins can be refocused using an NMR pulse sequence during storage \cite{non_ideal_polarization, albrecht2015controlled}.

A quadrupolar inhomogeneity of $\sigma(\Delta_Q) = 50$ kHz, achievable for GaAs/AlGaAs QDs \cite{chekhovich2017measurement}, will contribute a multiplicative factor of $F_\text{quad} = 0.996 $ to the transfer fidelity, for the $\Delta m = 2$ mode over the timescale of a full write-read cycle ($t_\text{trans} = 2\times 165$ ns).

The total state transfer fidelity, considering the mentioned contributions, is given by
\begin{equation}
    F_\text{transfer} = F_\text{e,init} \times F_\text{n,init} \times F_\text{quad}  .
\end{equation}
A state transfer fidelity of $F_\text{transfer} = 0.993$ ($F_\text{transfer} = 0.973$) for a full write-read cycle can be obtained for a nuclear polarization of $95\%$ ($80\%$), considering the values discussed in this section.

\subsection{Photon scattering}
\label{photon_scattering_fid}

The fidelity of the photon scattering scheme in the limit of $C=4g^2/(\kappa\gamma)\gg1$ is given by \cite{Faezeh-gate}:
\begin{equation}
\label{eq:kimble}
\begin{aligned}
F_\text{gate}&= 1-\frac{5}{2C}- \frac{\left(\delta_{\epsilon_1}\!-\delta_{\epsilon_2}\right)^2}{2\gamma^2C} -\xi T_\text{gate} \\&
-\frac{\sigma_p^2+\delta_p^2}{4 \gamma^2 C^2}\!\left[11 - 20\left(\frac{2g}{\kappa}\right)^2\! + 12 \left(\frac{2g}{\kappa} \right)^4\right]
\! ,
\end{aligned}
\end{equation}
where $T_\text{gate}=8\pi\sqrt{2\ln{2}}/\delta_p$ is the gate time defined to be twice the FWHM duration of the photon, $\delta_p$ is the photon's spectral standard deviation, $\sigma_p$ is the spectral diffusion standard deviation, and $\delta_{\epsilon_i}$ ($i\in\{1,2\}$) is the detuning of the $i$-th QD’s optical transition from the cavity resonance. The fidelity definition used here is the square of the definition used in Ref. \cite{Faezeh-gate}, and as such, we have modified Eqn. \ref{eq:kimble} to be consistent with the other definitions of fidelity used in our work. Eqn. \ref{eq:kimble} was derived by assuming that the incoming photon has a Gaussian spectral shape, however, photons emitted from a QD-based single-photon source are likely best described by a Lorentzian shape. We will approximate the incoming photon to have a Gaussian spectral shape, as it not expected to significantly affect the overall fidelity.

We limit our analysis to the high-performance limitations in our scheme by assuming that decoherence processes other than cavity dissipation and spontaneous emission occur on timescales that are much longer than the gate time. These additional decoherence processes are included in our analysis by a single effective decoherence rate, $\xi$, which we assume is dominated by the electron spin decoherence, $\xi = 1/2T_2$. In a high cooperativity regime the probability for the incident photon to excite either of the QDs is very low because the probability of the QD absorbing the photon is inversely proportional to the cavity cooperativity. Therefore, we neglect the optical pure dephasing for this scheme. Eqn. \ref{eq:kimble} is valid to first order in $\xi T_\text{gate}$, $C^{-1}$,  and $(\gamma T_\text{gate})^{-1}$; and to second order in $\delta_{\epsilon_k}/\gamma$, $\sigma_p/\gamma C$ and $\delta_p/\gamma C$.

The following QD parameters are used in the fidelity calculation $\gamma=2\pi \times 0.59$ GHz \cite{zhai2020low}, $\xi=1/2T_2 = 2\pi \times 1.6$ kHz where $T_2=50\,\mu$s (discussed in Sec. \ref{implementation_entanglement_generation}) is the electron spin coherence time, $\sigma_p=2\pi \times 500$ MHz, $\delta_{\epsilon_1}=\delta_{\epsilon_2}$, and $g/\kappa=0.1$. A source photon with $\delta_p = 2\pi \times 2.4$ GHz (corresponding to lifetime $1/\gamma$ and Purcell enhancement of $3$) gives $T_\text{gate}= 8\pi\sqrt{2\ln{2}} /\delta_p = 2$ ns. If we assume that the population lifetime is equal to the radiative lifetime, as we did in Sec. \ref{entanglement_gen_fid}, then $F_p = C\gamma/\gamma_r = C$. For a Purcell factor of $F_p=500$ ($F_p=200$) the gate fidelity is $F_\text{gate}=0.995$ ($F_\text{gate} = 0.986$).

\subsection{State measurement}
\label{Section_readout}

The fidelity of the state measurement depends on the branching ratio and the spin selectivity of the excitation. For readout in the pseudo-Faraday configuration, Ref. \cite{AC_Stark_proposal} estimated a fidelity of $F_\text{readout} = 0.762$ without the use of a coupled cavity, using feasible QD parameters for real experiments. To put this estimation into context, $F_\text{readout} = 0.823$ has been measured for InGaAs QDs in the true Faraday configuration \cite{delteil2014observation}.

In the pseudo-Faraday configuration the theoretical probability for a QD, excited through the transition $\ket{\downarrow} \rightarrow \ket{\uparrow \downarrow \Downarrow}$, to decay back to the initial ground state is $0.98$ \cite{AC_Stark_proposal}. This probability could be further increased to unity in the presence of a resonant cavity. Therefore,
repeated excitation of the QD through the $\ket{\downarrow} \leftrightarrow \ket{\uparrow\downarrow\Downarrow}$ cycling transition will emit a large number of photons into the cavity if the electron is in the $\ket{\downarrow}$ state and eventually one will be detected.

Exciting the QD for longer than the Purcell-enhanced lifetime is not an issue for state measurement as it is for entanglement generation because we are not required to minimize the possibility of multiple photon emission and, hence, the frequency bandwidth of the excitation laser can be sufficiently small to resolve the transition. It is therefore possible to perform the state measurement using continuous-wave excitation.

The rate of emission from the system under continuous-wave excitation is given by $\Omega^2/\gamma^\prime$, for a driving amplitude $\Omega \ll \gamma^\prime$. Assuming that dark counts and detection events follow the Poisson distribution, the state measurement fidelity is given by \cite{ji2020proposal}:
\begin{equation}
    F_\text{readout} = \frac{1}{2} \Bigg[ 1 + e^{-T D} - e^{\frac{-T \eta_c \eta_d \Omega^2}{{\gamma^\prime}}} \Bigg] ,
\end{equation}

where $T$ is the readout time and $D$ is the dark-count rate. We consider $D=500$ Hz \cite{eisaman2011invited}, $\eta_c = \eta_d = 0.9$, and a readout time of $T=600$ ns to maximize the fidelity. For a Purcell factor of $F_p = 500$ ($F_p = 200$) the state measurement fidelity is $F_\text{readout} = 0.99983$ ($F_\text{readout} = 0.99985$).

\subsection{Overall Fidelity}
\label{overall_fid}

The overall fidelity of the repeater protocol can be estimated by multiplying the fidelity associated with each of the components of our scheme. For a QD-based repeater with $l=2^n$ elementary links, the overall fidelity is given by
\begin{equation}
\label{overal_fidelity_eqn}
    \begin{aligned} 
        F_\text{total} =  (F_\text{e,init})^{2l} &\times (F_\text{readout})^{2(l-1)} \\
         \times(F_\text{ent} &\times F_\text{transfer}^2)^l \times (F_\text{gate})^{l-1} ,
    \end{aligned} 
\end{equation}
which is accurate only in the high-fidelity regime.

\begin{figure}
	\centering
	\includegraphics[trim=0cm 0cm 1cm 1cm, clip=true, width=8.6cm]{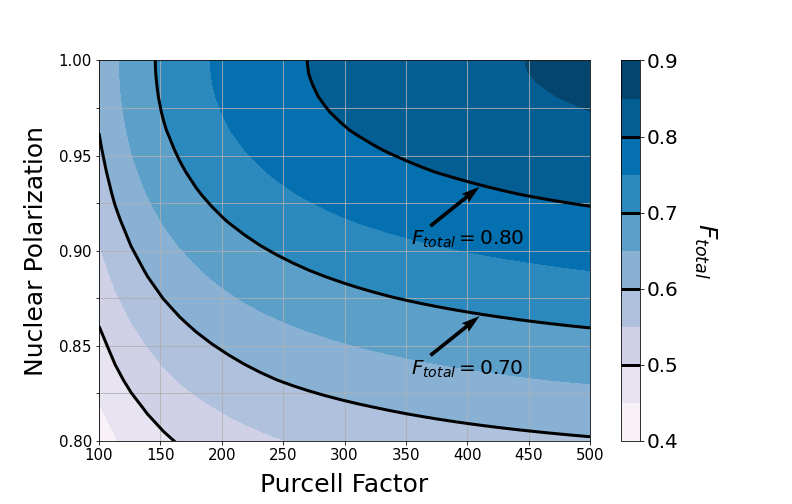}
	\caption{Contour plot of the QD repeater overall fidelity as a function of the Purcell factor and nuclear polarization. We consider a repeater with a nesting level of $n=3$ and the same values used in the last three subsections.}
	\label{fig:overall_fidelity}
\end{figure}

In Fig. \ref{fig:overall_fidelity} we present a contour plot of the overall fidelity as a function of the Purcell factor and the nuclear polarization. We consider $8$ local links, $F_p = C\gamma/\gamma_r = C$, $\kappa = 2\pi \times 100$ GHz, a QD-cavity detuning of $2\pi \times 275$ GHz during the entanglement generation procedure, and the same values used in the last three subsections.

For a nuclear polarization of $95\%$ and $F_p=500$ ($F_p=200$) our quantum repeater will have an overall fidelity of $F_\text{total}=0.831$ ($F_\text{total}=0.734$). A polarization of $80\%$ and $F_p=500$ ($F_p=200$) results in an overall fidelity of $F_\text{total}=0.596$ ($F_\text{total}=0.526$). In the idealized case where the nuclear polarization is $99.9\%$, the overall fidelity is still limited to $F_\text{total}=0.858$ ($F_\text{total}=0.758$) for $F_p=500$ ($F_p=200$). Thus, a significant improvement to the overall fidelity is not obtained when considering a nuclear polarization above $\sim95\%$.

The overall fidelity can be further increased by considering the use of higher cooperativity cavities to improve the gate fidelity, and by reducing the spectral diffusion to improve the entanglement generation fidelity. The use of etalon filters would also improve the fidelity of both the gate and the entanglement generation steps, albeit at the expense of reduced efficiency. It may also be possible to increase the fidelity by adapting our scheme to include entanglement purification, in which many entangled pairs are simultaneously generated and then combined into a smaller number of entangled pairs with lower errors by performing two-qubit gates \cite{dur1999quantum, kalb2017entanglement}.

\section{Implementation}
\label{Implementation}

In order to be advantageous, our scheme requires that the nanofabrication techniques for low-strain QDs advance to the point where extraction efficiencies of a least 50\% are achievable. In addition to high extraction efficiencies we require the ability to control the nuclear spins to the precision outlined in Sec. \ref{Fidelity}, Purcell factors in excess of 100, and the ability to integrate two QDs in the same cavity. Integration of GaAs/AlGaAs QDs with photonic nanostructures is still in its infancy, and as such, we believe that this is currently the largest roadblock towards achieving our repeater scheme.

Encouragingly, recent experimental demonstrations have shown that low-strain GaAs/AlGaAs QDs can be incorporated with circular Bragg grating \cite{liu2019solid}, optical antenna \cite{chen2018highly}, and photonic crystal (PhC) cavities \cite{wei2020bright}. The latter is particularly noteworthy, as in the next subsection we propose the use of PhC cavities for our scheme. Ref. \cite{wei2020bright} reported an extraction efficiency of $\sim 23\%$, however, no Purcell enhancement was obtained.

The recent interest in GaAs/AlGaAs QDs can be attributed to research of the biexciton-exciton radiative cascade which can generate polarization-entangled photons. Recent work \cite{scholl2020crux} has revealed that the time jitter of the cascade limits photon indistinguishability to $M = 66\%$, unless the ratio of the biexciton lifetime to the exciton lifetime is made very small through cavity engineering. This, in turn, limits the entanglement generation fidelity to $(1+M)/2 = 0.83$ per Bell-state measurement. As such, efforts in this field will likely push for advances in microcavity fabrication methods compatible with GaAs/AlGaAs QDs, which would also prove beneficial to the implementation of our protocol. Furthermore, if the limited indistinguishability ultimately prohibits the use of these photon pairs in a successful repeater protocol, then our work highlights the fact that there are other promising applications of GaAs/AlGaAs QDs for quantum communication.

Our scheme has additional demands which include the ability to obtain levels of nuclear spin polarization greater than $80\%$, improved electron spin and nuclear spin coherence times, reduced levels of spectral diffusion, and efficient frequency conversion techniques. In this section we discuss the state-of-the-art and outlook for several of the figures of merit required by our scheme.

\subsection{QD-cavity system}
\label{implementation_cavity}

As previously mentioned, we propose the use of NFDE GaAs/AlGaAs QDs \cite{atkinson2012independent}. The primary reason for selecting this type of QD is to mitigate the effect that inhomogeneous quadrupolar splittings have on the state transfer fidelity.

We consider the use of planar PhC cavities due to their ease of fabrication using standard etching and lithography techniques \cite{on_chip_phcs}. Well-developed techniques allow for the high-precision fabrication of PhC cavities and QDs can be naturally incorporated \cite{lodahl2015interfacing}. The cavities allow for an embedded QD to be probed by laser excitation from above the cavity. An additional motivation for considering the use of PhC cavities can be seen in the requirement that the two QDs located in each node are isolated from the neighboring QDs in the sample. While it is possible to mask these neighboring dots, such an approach is typically incompatible when considering the use of microcavities or waveguides as the masking process can modify the optical properties of these structures. When considering PhC cavities, the cavity hole patterns can be designed such that the etching process removes all the QDs, except for the two dots located in the cavity. Such an approach was utilized in the work of Ref. \cite{calic2017deterministic}, where the deterministic integration of two InGaAs QDs with a planar linear three-hole defect (L3) PhC cavity was demonstrated for two QDs separated by $350$ nm.

PhC cavities with high $Q$-factors and small mode volumes are potentially advantageous for achieving large Purcell enhancements \cite{kress2005manipulation}. It has been suggested that $F_p=43$, obtained via resonant excitation of InGaAs QDs, could be increased to $F_p\sim 200$ by increasing the $Q$-factor by a factor of 5 \cite{liu2018high}. Surface passivation techniques have been shown to significantly increase $Q$-factors in coupled QD-cavity systems \cite{kuruma2020surface, najer2019gated}. The combination of resonant excitation and surface passivation techniques represent a promising path towards achieving Purcell enhancements in excess of 100.

An important distinction between InAs/GaAs and GaAs/AlGaAs QDs can be seen in the roughness of the AlGaAs membrane, which may limit achievable $Q$-factors. Additionally, Al is prone to oxidation when exposed to air. A possible route for overcoming both of these issues may be to use AlGaAs layers with lower Al-concentration, as in Ref. \cite{zhai2020low}.

Our scheme also requires the use of waveguides for directed photon propagation in both the CAPS gate and entanglement generation steps. We propose the use of PhC waveguides as they can be naturally incorporated with PhC cavities and offer a scalable platform which could lead to on-chip implementations \cite{englund2007generation}. The emission of photons into a desired waveguide with a probability greater than $98\%$ has been reported for InAs QDs embedded in PhC waveguides \cite{arcari2014near} and coupling efficiencies of up to $90\%$ have been obtained using a PhC cavity-waveguide system \cite{faraon2007efficient}.

\subsection{Entanglement Generation}
\label{implementation_entanglement_generation}

Entanglement generation requires excitation of the $\ket{\downarrow} \rightarrow \ket{\uparrow\downarrow\Downarrow}$ transition. To excite only this transition, the spectral width of the laser pulse should be less than the ground and excited state splittings. When the nuclear spins are polarized, the ground and excited state splittings are determined by the strength of both the external magnetic field and the Overhauser field. Considering the case where the nuclear polarization increases the splitting, the electron ground levels are split by $\Delta E_g = \Delta E_Z^g  + \Delta E_\text{OH}$ \cite{Nuclear_spins_QDs}, where $\Delta E_Z^g = |\mu_B g_e B_x| / h$, $\Delta E_\text{OH}$ is the Overhauser shift, $\mu_B$ is the Bohr magneton, $g_e$ is the electron $g$-factor, and $B_x$ is the applied magnetic field. Similarly, the excited (trion) levels are split by $\Delta E_e = |\mu_B g_h B_x| / h  + \Delta E_\text{OH}$, where $g_h$ is the hole $g$-factor.
	
The maximum possible Overhauser shift in GaAs is $\Delta E_\text{max}^\text{OH} \approx 31 $ GHz ($127 \,\mu$eV) \cite{ulhaq2016vanishing, chekhovich2017measurement}. For the values $B_x = 6.6$ T, $g_e = -0.076$, $g_h = 1.309$ \cite{zhai2020low}, and a nuclear polarization of $80\%$ ($\Delta E_\text{OH} = 25$ GHz), we have $\Delta E_e = 146$ GHz and $\Delta E_g = 32$ GHz, which provides an upper limit for the spectral width of the entanglement generating laser pulse.

The transform-limited duration for a Gaussian pulse with a spectral width of $\Delta \nu = 32$ GHz is $\Delta t= 14$ ps. For $F_p>18$ and $\gamma_\text{r}= 2\pi \times 0.59$ GHz, the Purcell-enhanced lifetime is less than $14$ ps. Therefore, for Purcell factors in excess of 18 the entanglement generation pulse could result in the re-excitation of the QD and the emission of multiple photons. As such, each QD must be detuned from the corresponding cavity during the entanglement generation step to reduce the effective Purcell effect.

The electron spin coherence time used in the fidelity calculations assumes that $T_2 = 50\,\mu$s (for $B=6.6$ T). The dominant mechanism for electron spin decoherence is predicted to be the precession of the electron spin in the fluctuating Overhauser field \cite{merkulov2002electron}, as relaxation due to phonons at low temperatures can be neglected. In the absence of nuclear spins it is predicted that the spin coherence should approach the ideal limit $T_2 = 2 T_1$ \cite{golovach2004phonon}.

Fluctuations in the Overhauser field are a result of both electron-mediated nuclear spin flips (either direct or virtual) and nuclear dipole-dipole interactions. Electron-mediated decoherence can be minimized by considering the combination of a large magnetic field and a high nuclear spin polarization \cite{deng2006analytical}. As previously mentioned, nuclear spin fluctuations are significantly reduced in NFDE QDs \cite{ulhaq2016vanishing}. Inhomogeneous quadrupolar splittings lift the degeneracy between nuclear spin states, reducing the rate of dipole-mediated nuclear spin diffusion. We believe that the combination of a large magnetic field, high levels of nuclear polarization, and the unique levels of strain in NDFE QDs can lead to electron spin coherence times approaching the ideal limit of $2T_1 = 98 \,\mu$s, for $T_1 = 48\,\mu$s \cite{zhai2020low}. Additional methods to extend the electron spin $T_2$ include dynamical decoupling techniques \cite{clark2009ultrafast, viola1999dynamical, khodjasteh2005fault}, or driving the electron spin to modify the properties of the nuclear bath \cite{ethier2017improving, stepanenko2006enhancement, greilich2007nuclei, reilly2008suppressing, vink2009locking}.

Entanglement generation also requires high interference visibility \cite{huber2015optimal}. Single photons have been generated on-demand from an InGaAs QD with an end-to-end efficiency of 57\% and with a two-photon interference visibility of 97\% that remains stable over a large time separation between photons \cite{tomm2020bright}. Particularly noteworthy to our work is the recent demonstration in which a two-photon interference visibility of 93\% was achieved between spatially separated droplet-etched GaAs/AlGaAs QDs without the need for cavity enhancements, temporal post-selection, spectral filtering, or active frequency stabilization \cite{zhai2021quantum}.

To obtain highly indistinguishable photons, spectral diffusion of the optical transition needs to be controlled such that the line broadening is close to the transform limit, that is, the spectral bandwidths of the interfering photons are limited only by their radiative lifetime \cite{houel2012probing}. The dynamic environment of a QD leads to variations in the optical transition energy through several decoherence processes including phonon-induced dephasing \cite{tighineanu2018phonon}, nuclear spin noise, and charge fluctuations \cite{vural2020perspective}. The work of Ref. \cite{zhai2020low} has demonstrated that improved charge control alongside reduced Al-concentrations in the AlGaAs membrane can lead to optical linewidths close to the lifetime limit. We expect that future efforts using high-quality nanofabrication techniques, nuclear ensemble spin squeezing, and charge-controlled QDs will limit the amount of spectral diffusion to a few hundred MHz, resulting in the high-fidelity implementation of our scheme as discussed in Sec. \ref{Fidelity}.


\subsection{Entanglement swapping}
\label{Ph-scattering-Imp}
The derivation of Eq. \ref{eq:kimble} in Ref. \cite{Faezeh-gate} assumed that both of the detunings $\delta_{\epsilon_1}$ and $\delta_{\epsilon_2}$ were small compared to the optical decay rate $\gamma$. Therefore, to get the upper bound on these detunings individually, the next higher order perturbation terms should be considered. Here we assume that the QDs are tuned such that their resonant frequencies are equal, yet there still will be some infidelity over time due to the spectral diffusion of the QDs inside the cavity. We estimate that the amount of the spectral diffusion that can be tolerated is several hundred MHz. Note that any amount of detuning between the optical transition frequencies of the QDs can cause infidelity. Spectral diffusion greater than a few hundred MHz will cause infidelity more than the infidelity caused due to the finite cavity cooperativity.

An alternative method to implement a controlled phase gate would be to use the electric dipole-dipole interaction between trions in neighboring QDs \cite{Simon_qds, asadi2020protocols}. The advantage of a gate based on the dipole-dipole interaction is that the gate is deterministic which is not the case for CAPS-based gates. As mentioned, the dipole-dipole interaction requires that the QDs be spaced relatively close together and hence, spectral addressing of the dots would be required in the entanglement generation step. The CAPS method, on the other hand, would allow for either spectral or spatial addressing of the dots.

It is also possible to perform a deterministic gate via the exchange of a virtual cavity photon \cite{Faezeh-gate}. In this scheme one transition of the QDs should be in resonance while they are dispersively coupled to a cavity mode. Note that in the derivation of the gate fidelity in Ref. \cite{Faezeh-gate} it has been assumed that $\delta_{eg}\gg 2\pi T_o^{-1}$ where $T_o$ is the optimal gate time and $\delta_{eg}=|\Delta E_e-\Delta E_g|$. This condition is violated in our system because $T_o$ is inversely proportional to the relatively large optical decay rate associated with the QD trion. Thus, the fidelity of this scheme, which is dictated by the QD-cavity detuning, should be optimized by selecting an appropriate detuning.

A gate performed using the exchange of a virtual cavity photon is well-suited for systems with a small amount of optical pure dephasing. In contrast, the CAPS method does not require excitation and is therefore of interest to systems with a larger amount of optical pure dephasing. Another advantage of the CAPS method is that the associated fidelity scales as $1-5/4C$, whereas the fidelity of the virtual photon exchange scheme scales as $1-\pi/\sqrt{C}$. The fidelity of each of these schemes as a function of the cavity cooperativity is illustrated in Fig. 7 of Ref. \cite{Faezeh-gate}.

\subsection{Frequency  Conversion}
\label{freq_con}
The entanglement generation rates calculated in Sec. \ref{Rates} considered the efficient conversion of the $\sim 750$ nm photons emitted from the QDs to the telecom wavelength range.  One option to perform the conversion is to use a method of difference frequency generation in which the emitted photons are mixed with a pump pulse at $1450$ nm in a periodically poled lithium niobate (PPLN) waveguide \cite{de2012quantum}. The PPLN waveguide efficiently converts the $750$ nm photons to $1550$ nm photons, conditional on overlap with the pump pulses. After narrow-band filtering to eliminate residual scattered light, detection of $1550$ nm photon would herald the detection of a single $750$ nm photon.

External conversion efficiencies of up to  40\% have been obtained using PPLN crystals \cite{ates2012two, zaske2012visible}, and could be further increased by improving the coupling in and out of the crystal. The conversion loss enters the entanglement distribution rate quadratically. For a conversion efficiency of 40\%, the distribution rates of our scheme are $16\%$ of the results presented in Sec. \ref{Rates}. On-going research includes frequency conversion while retaining a high degree of single-photon indistinguishability \cite{kambs2016low}, and the development of scalable, on-chip frequency conversion devices \cite{singh2019quantum}.

It may also be possible to instead use QDs which emit at telecom wavelengths. For instance, recent work has reported the site-controlled growth of indium-arsenide-phosphide (InAsP) QDs embedded in a nanowire waveguide which emit at telecom wavelengths \cite{haffouz2018bright}. Continued research in this area may provide the option to use telecom QDs in our quantum repeater scheme.

\section{Conclusion}\label{Conclusion}

Recent work has experimentally verified that it is possible for the state of a QD electron-spin qubit to be reversibly transferred to the surrounding nuclear spins. Long-lived spin coherence of the QD nuclei make the nuclear ensemble a promising candidate to serve as a quantum memory in quantum information applications. Here, we proposed a quantum repeater based on single QD electron-spin qubits and nuclear-spin-ensemble memories.

Our scheme exploits several advantages of singly-charged QDs, namely the fast initialization and manipulation of electron-spin states, and the dedicated local memory per spin qubit offered by the nuclear ensemble. We draw inspiration from well-developed fabrication technologies for self-assembled QDs, and propose that GaAs/AlGaAs QDs are integrated with photonic nanostructures, allowing for the use of cavity-assisted photon scattering gates. We have shown that the achievable efficiency of our scheme could be higher than that of a scheme based on photon-pair sources and external memories. Currently, the largest obstacles to realizing our repeater scheme include the ability to control the nuclear spins, and the nanofabrication methods available for low-strain QDs. We require nuclear polarizations in excess of $80\%$, extraction efficiencies that are comparable to the state-of-the-art for self-assembled QDs ($\sim50\%$), Purcell factors exceeding 100, and the ability to integrate two QDs in a cavity.

The significant interest in the biexciton-exciton cascade, combined with the need integrate these QDs in cavities to overcome indistinguishability limitations, could lead to the rapid development of fabrication methods applicable to GaAs/AlGaAs QDs. Such advances would benefit the use of these dots as both photon-pair and spin-photon entanglement sources.

Finally, it is important to mention that there have been several recent quantum repeater proposals that avoid the use of quantum memories \cite{munro2015inside, azuma2015all}. These approaches typically achieve higher entanglement generation rates, but also require significantly more resources \cite{hilaire2021resource}. Extending upon the analysis of Ref. \cite{hilaire2021resource}, an interesting future direction could be to develop a hybrid QD-based repeater scheme which utilizes some combination of all-photonic and memory-based protocols.

\section*{Acknowledgments}
We thank Dan Dalacu, E.V. Denning, Olivier Krebs, Loic Lanco and Jordan Smith for useful discussions. This work was supported by the Natural Sciences and Engineering Research Council of Canada (NSERC) through its Discovery Grant, Canadian Graduate Scholarships, CREATE, and Strategic Project Grant programs; and by Alberta Innovates Technology Futures (AITF) Graduate Student Scholarship program.

\bibliographystyle{apsrev4-1}
\bibliography{ref}



\end{document}